\documentclass[sigconf, nonacm]{acmart}

\usepackage{graphicx}
\usepackage{subcaption} 
\usepackage{float}

\AtBeginDocument{%
  }

\begin{document}
\title{"See What I Imagine, Imagine What I See”: Human-AI Co-Creation System for 360° Panoramic Video Generation in VR}

\author{Yunge Wen}
\email{yw3776@nyu.edu}
\affiliation{
  \institution{New York University}
  \state{New York}
  \country{USA}
}
\begin{abstract}
The emerging field of panoramic video generation from text and image prompts unlocks new creative possibilities in virtual reality (VR), addressing the limitations of current immersive experiences, which are constrained by pre-designed environments that restrict user creativity. To advance this frontier, we present Imagine360, a proof-of-concept prototype that integrates co-creation principles with AI agents. This system enables refined speech-based text prompts, egocentric perspective adjustments, and real-time customization of virtual surroundings based on user perception and intent. An eight-participant pilot study comparing non-AI and linear AI-driven workflows demonstrates that Imagine360's co-creative approach effectively integrates temporal and spatial creative controls. This introduces a transformative VR paradigm, allowing users to seamlessly transition between 'seeing' and 'imagining,' thereby shaping virtual reality through the creations of their minds.
\end{abstract}

\begin{teaserfigure}
  \includegraphics[width=\textwidth]{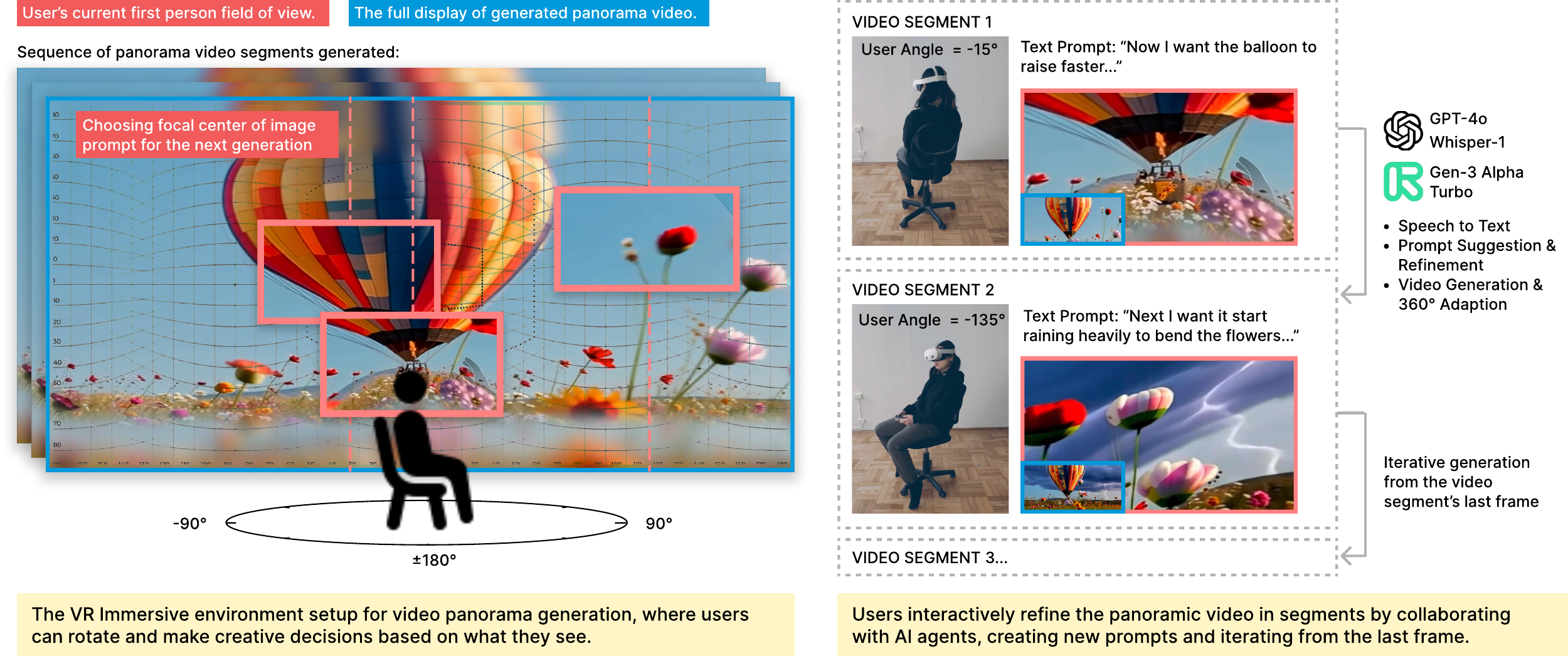}
  \caption{The boundary between experience and representation has long been debated in philosophical inquiry. Building on the recent advancement of panorama video generation, we present \textbf{Imagine360}—a system that enables users to co-create with AI agents, freely transforming their surroundings in virtual reality and reshaping time and space according to their will, making perception a malleable construct in the context of modern philosophical exploration. For example, a user imagines the serene weather turning into a thunderstorm and iteratively envision the next segment based on what they see.}
\end{teaserfigure}

\maketitle

\section{Introduction}
Can we shape our reality through the creations of our minds? Throughout history, philosophers have debated the relationship between reality and perception. Kant argued that we cannot directly access the noumena—the objective reality behind phenomena—through our senses, leaving our experience ultimately confined to the world of appearances \cite{kant1790critique}. Schopenhauer, on the other hand, viewed the world as a representation of the will, shaped by our senses and cognition to reflect its nature \cite{schopenhauer1818world}. In contrast, Eastern philosophy transcends this dualism, as illustrated by the "Butterfly Dream": when I dream of being a butterfly, it might just as well be the butterfly dreaming of being me, suggesting the fluidity and interdependency of self and reality \cite{butterfly}. Today, with advancements in virtual reality (VR) and generative AI, we may overturn past theories and make virtual reality a new means of expressing our being. As envisioned in science fiction \cite{sf_gibson1984neuromancer, sf_nolan2010inception, sf_stephenson1992snowcrash, sf_wachowskis1999matrix}, VR could enable the creation of immersive, self-generated realities that transcend objective physical representation.

Despite this compelling promise, most immersive environments today remain pre-designed, relying heavily on pre-recorded footage, CGI-rendered scenes, or 3D reconstruction techniques \cite{tochilkin2024triposr, dalal2024gaussian}. While early attempts have explored interactive VR painting \cite{vrpainting, vrpaintingfinger, immersivesketch} to enhance user creativity, the integration of real-time GenAI systems for co-creation remains underexplored. Building on video generation as a new artistic medium \cite{animatediff}, the state of the art \textit{360DVD} by Q. Wang et al. \cite{360dvd} has demonstrated the creative potential of generating 360° panoramic videos from text and image prompts. These videos allow users to explore scenes freely from any angle through equirectangular projection, which is characterized by (1) a 2:1 aspect ratio, (2) continuous left and right edges corresponding to the same meridian, and (3) motion patterns that often follow curved trajectories \cite{panoramat2i}, enabling greater flexibility in generating spatial and temporal dynamics compared to current systems.

To further explore this potential, we propose \textbf{Imagine360}, a proof-of-concept prototype that integrates co-creation principles \cite{machineintheloop, apprentice, boden2004creative} with AI agents for panoramic video generation in immersive environments. Our system enables users to generate panoramic videos with AI assistance, evaluate generation outcomes in real time, provide speech-based text prompts that are refined and supported by the AI agent, and recenter the panorama video’s focal point based on an egocentric perspective. Through a user study, we compared the human-agent co-creation strategy employed in our system to non-AI and linear AI-driven design scenarios, validating the effectiveness of our system. In conclusion, this work proposes a new paradigm for co-creative VR experiences, where users can "see what I imagine and imagine what I see."

\section{Related Works}

\subsection{The Emerging Technology of Panoramic Video Generation }

The Denoising Diffusion Probabilistic Model (DDPM) \cite{ddpm_cheng2023null, ddpm_ho2020denoising} has demonstrated exceptional success in generating high-quality images, while text-to-image (T2I) diffusion models \cite{t2i_mou2024t2i, t2i_akimoto2022diverse, t2i_nichol2021glide} showcase remarkable capabilities in creating images from user-provided prompts. These advancements in image generation have naturally extended to text-to-video (T2V) generation \cite{animatediff, latentshift, makeavideo, magicvideo}, leveraging space-time separable architectures that inherit spatial operations from pre-trained T2I models, significantly reducing the complexity of constructing space-time models from scratch. While GAN-based methods for generating panoramic images have been extensively explored, research on panoramic video generation remains underexplored. A notable breakthrough came in 2024 with 360DVD \cite{360dvd}, which introduced an innovative pipeline for panoramic video generation. This approach integrated a lightweight 360-Adapter and sliding window techniques to adapt pre-trained AnimateDiff models for panoramic content. Despite its advancements, 360DVD still struggles to produce complex and diverse motion patterns compared to conventional 2D video generation. Our study combines conventional video generation methods with alternative video processing techniques to unlock the potential of panoramic video applications in VR.

\subsection{Human-AI Co-Creation in Immersive Environment}

Creativity is a fundamental driver of innovation, and the introduction of AI agents presents transformative opportunities. Many human-computer interaction theories have explored this synergy. The 'machine-in-the-loop' framework \cite{machineintheloop} emphasizes human control with AI serving as a supportive tool, while the Apprentice Framework \cite{apprentice} delineates distinct human-AI roles for collaborative functions. Similarly, Kantosalo et al. \cite{kantosalo2016creative} propose a dynamic collaboration model in which humans and AI alternate tasks to achieve shared creative goals. Building on these theories, existing generative AI tools for co-creativity predominantly focus on text and image generation \cite{ibarrola2023collaborative, chung2022talebrush, gero2022sparks, chilton2021visifit, gmeiner2023exploring} but have limited applications in immersive environments. Leveraging VR painting as a creative medium \cite{vrpainting, vrpaintingfinger}, \textit{ImmerseSketch} \cite{immersivesketch} uses diffusion models and depth estimation to transform 2D prompts into immersive 3D environments. \textit{Interact360} \cite{interact360} integrates generative AI into VR to generate user portraits and blend them seamlessly into panoramic scenes. Building on these developments, our work integrates AI agents seamlessly into the environment control system, operating invisibly to refine workflows and provide background suggestions. This approach preserves the user's sense of full control and freedom, fostering a connection between will and perception to create a self-curated reality.

\begin{figure*}[h]
    \centering
        \includegraphics[width=\textwidth]{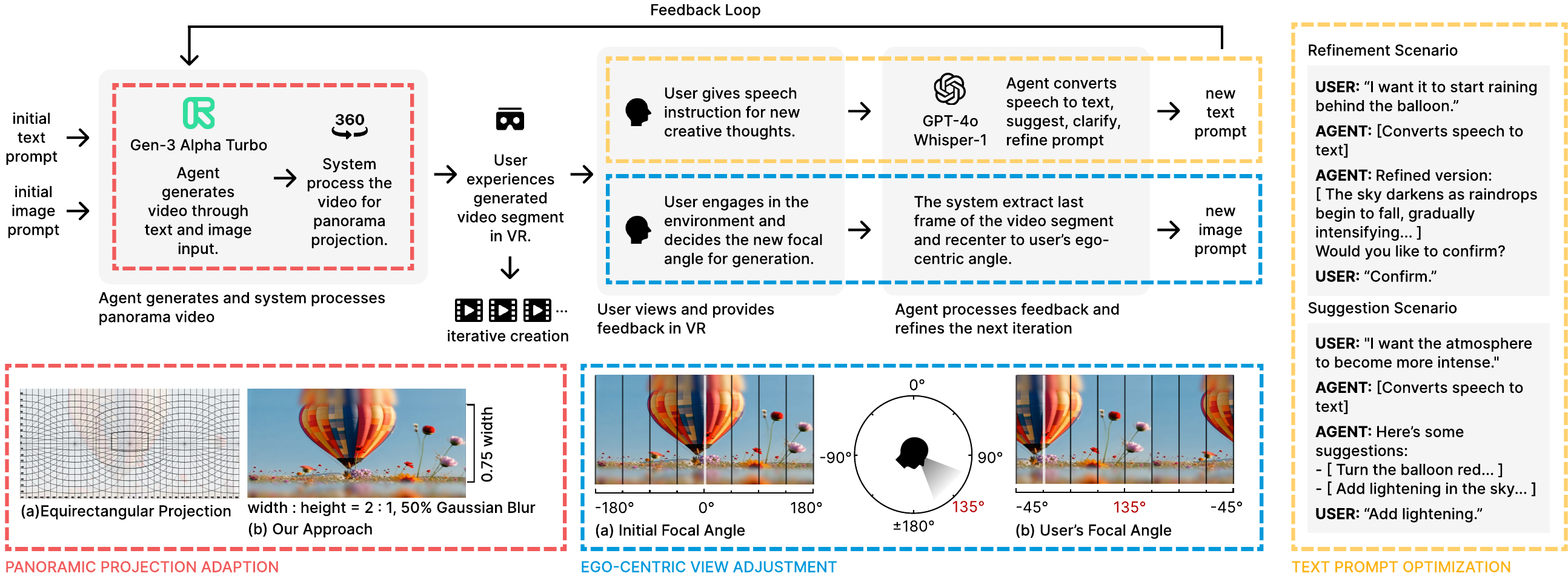}
        \caption{\textbf{Co-Creation Workflow of Panorama Video Generation in VR.} The system enables users to (1) generate panoramic videos with AI assistance, (2) evaluate generation outcomes in real time, (3) provide speech-based text prompts that are refined and supported by the AI agent, and (4) recenter the panorama video’s focal point based on an egocentric perspective. This co-creation workflow leverages AI's capabilities in the backend and offers suggestions upon inquiry that seamlessly connect the user's intent with their perception.}
    
\end{figure*}

\section{System Design}

Imagine360 is a proof-of-concept prototype that enables users to interactively imagine, prompt, and control the temporal and spatial dimensions of panorama video generation, leveraging AI collaboration to reflect their instant inspiration based on immediate perception. The system consists of three core components: video generation, panorama projection, and interaction with AI agents.

\textbf{Video Generation:} Given the lack of an optimal model for panorama video generation, the system leverages a conventional 2D video generation approach, utilizing the Runway Gen-3 Alpha Turbo API image-to-video model. To tailor outputs for panoramic scenarios, panorama-specific descriptors are appended to the text prompts. This approach ensures smooth motion dynamics while preserving the immersive VR perspective.

\textbf{Panorama Projection:} The generated 2D videos are transformed to fit an equirectangular projection format through post-processing. The aspect ratio is adjusted to achieve a 2:1 output, while edge blending is applied to ensure visual continuity. The background is blurred using a 50\% Gaussian filter, and the height of the foreground is reduced to 75\% of the original to enhance visual coherence. The processed videos are integrated into a VR environment using Unity and Python socket communication, with an alternative solution that the final outputs imported into the Skybox application via a data connection.

\textbf{Interaction with the AI Agent:} Following the initial video viewing, users can interact with the system to generate subsequent videos by issuing voice commands and specifying adjustments to their visual focus:

\begin{enumerate}
    \item \textbf{Voice Input and Text Prompt Optimization:} User voice inputs are processed using OpenAI’s Whisper-1 model to generate text prompts. These prompts are then optimized using GPT-3.5 Turbo to ensure alignment with user intentions, enhancing the quality and relevance of the generated video outputs.
    \item \textbf{Ego-centric View Adjustment:} Users can redefine the image prompt by selecting a new visual center. The system maps image length to angular degrees. By default, the last frame of the previously generated video serves as the initial image prompt (0°). However, users may specify a preferred focal direction (e.g., +45° or -90°), enabling the system to adjust the 360-degree projection to align with their chosen visual to enhance creative control.
\end{enumerate}

\textbf{Co-Creation Workflow:} The workflow begins with the user providing an initial text or image prompt, which the agent uses to generate a panoramic video. The generated video is processed and experienced by the user in a virtual reality environment. During this process, the user can provide feedback through speech instructions or by adjusting their ego-centric view. The agent processes this feedback by converting speech to text, refining the prompts, extracting the last frame, or re-centering to a user-chosen angle. This feedback loop allows the agent to iteratively enhance the video generation process, enabling the creation of more meaningful and tailored outputs in the next iteration.

\section{Pilot Study}

To validate the effectiveness of our system and gather design insights, we conducted a pilot study comparing our approach with non-AI and linear AI-driven workflows.

\subsection{Parameters and Procedure}
\textbf{Task 1: Generic 360° Video (Baseline)}: In the baseline task, participants experienced non-generative AI content consisting of three pre-recorded 360° video clips, each lasting 30 seconds and representing the following categories: (1) \textbf{Outdoor}: A real-life skiing scene captured with a 360° camera, (2) \textbf{Indoor}: A virtual tour of a British palace, and (3) \textbf{Imaginary}: A computer-simulated journey falling into a black hole.

\textbf{Task 2: Linear AI-Driven Design}: In this task, participants receive the generation results directly without interacting with the agent. Participants were instructed to prepare three text prompts, one for each category (Outdoor, Indoor, Imaginary), and to supply corresponding images sourced from personal photos, online content, or AI-generated outputs.

\textbf{Task 3: Human-Agent Co-Creation}: In Task 3, we employed a Wizard of Oz (WoZ) approach, manually importing the generated videos into the VR headset and adjust user’s visual center. This was due to the following reasons: (1) the video generation API was not easily integrated with Unity, and Python-Unity communication was inconsistent; (2) the video generation process had variable durations, which could affect the experimental results. The WoZ method allowed us to simulate the system's intended functionalities, and these issues will be addressed and improved in future studies.

This task involved an iterative co-creation process in which participants collaborated with an AI agent to create a 30-second immersive video. The task was divided into three 10-second video segments:

\begin{itemize}
    \item \textbf{Segment 1}: Participants provided an initial text and image prompt, which was used to generate the first 10-second segment.
    \item \textbf{Segment 2}: After viewing the initial segment, participants adjusted their visual center using a rotating chair and either reused the original prompt or provided a new one to guide the generation of the next segment. The AI agent assisted by refining the prompts, incorporating panorama-specific descriptors to better align the output with user intentions.
    \item \textbf{Segment 3}: This iterative process was repeated to generate the final 10-second segment.
\end{itemize}
The three segments were combined into a single 30-second video for participants to review. 

Eight participants (3 male, 5 female), aged 18–30 years (mean = 25.5, SD = 2.39), were recruited from diverse professional fields, including design, media, machine learning, software development, law, and economics. Two had no design experience, four were amateurs or hobbyists, and two were professionals. One participant had extensive AR/VR interaction experience (5–30 hours), three had prior exposure to generative AI models (e.g., Stable Diffusion, MidJourney, DALL·E), and the rest had little to no experience with these technologies.

The experiment used a Meta Quest 3S VR headset. Participants sat in a rotating chair for free perspective adjustment and were guided to optimize their seating and headset fit for comfort.

Participants completed two assessments to evaluate cognitive workload and creativity. The NASA TLX \cite{tlx} measured workload across six dimensions on a 10-point Likert scale, while Boden’s Creativity Framework \cite{boden2004creative} assessed novelty, value, surprise, and relevance on a 7-point scale. After the tasks and surveys, participants joined a 10-minute semi-structured exit interview for qualitative feedback.

\begin{figure}[htbp] 
    \centering
    \begin{subfigure}[b]{\columnwidth}
        \centering
        \includegraphics[width=\columnwidth]{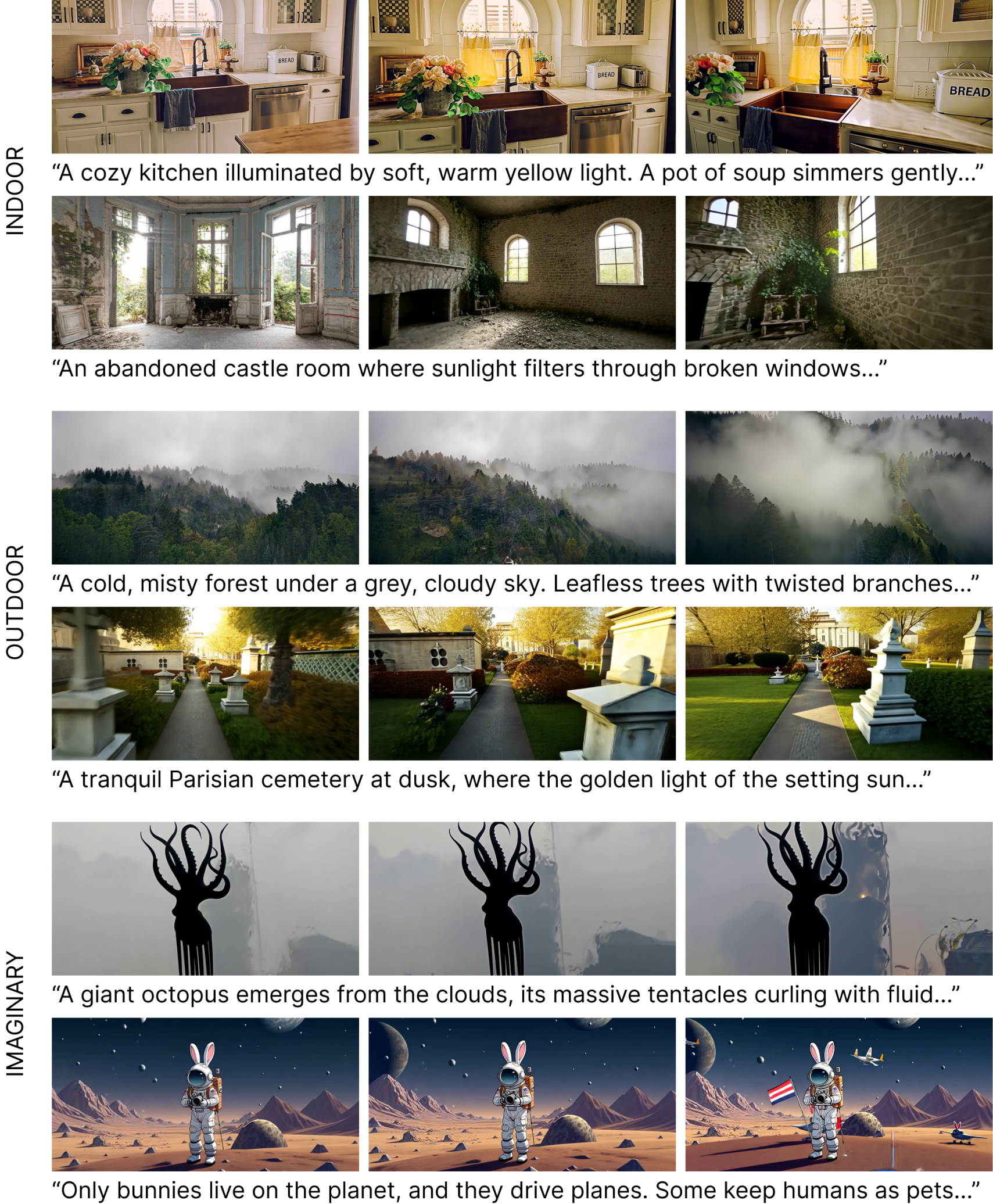}
        \caption{Linear AI-Driven Design}
        
    \end{subfigure}
    
    \vspace{1em} 
    
    \begin{subfigure}[b]{\columnwidth}
        \centering
        \includegraphics[width=\columnwidth]{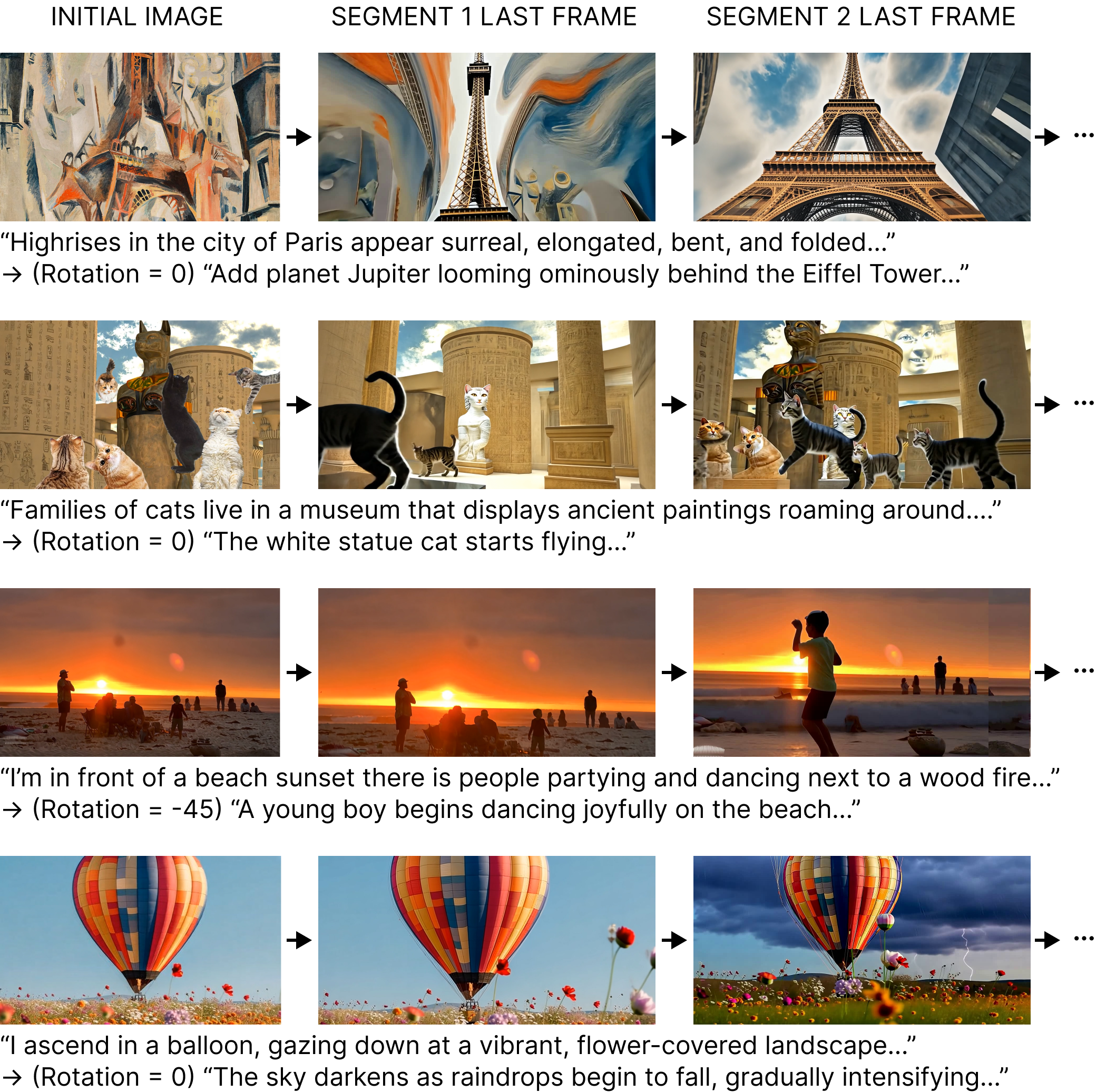}        \caption{Human-Agent Co-Creation}
    \end{subfigure}
    \caption{\textbf{Qualitative Comparisons of Participant Generation Outcomes.} Participants reported a preference for (b) iteratively collaborating with agents to refine generation outcomes and enhance creative potential, compared to (a) passively receiving generation results without agent interaction.}
\end{figure}

\subsection{Results}

To evaluate the effectiveness of our system, we analyzed quantitative metrics, observational insights, and interview data collected during the pilot study.

The co-creation framework achieved higher \textit{performance} but required greater \textit{mental} load. \textit{Performance} ratings were highest for co-creation (\textit{mean} = 8.875, \textit{SD} = 1.126), exceeding non-AI (\textit{mean} = 8.75, \textit{SD} = 2.435) and linear AI-driven design (\textit{mean} = 8.5, \textit{SD} = 1.604). A Friedman test showed significant effects on \textit{mental} ($\chi^2(3, N) = 7.36, p = 0.025$) and \textit{effort} ($\chi^2(3, N) = 7.66, p = 0.022$), with post-hoc Wilcoxon tests revealing higher \textit{mental} load for co-creation compared to non-AI ($p = 0.016$, corrected $p = 0.047$).

Our system shows significant \textit{relevance} and \textit{value} in outcomes and participants' creativity, with a strong positive correlation between the two (\( r = 0.81 \)). Co-creation exhibited a broader range of \textit{relevance} scores (3.0 to 6.5). Linear design scored higher in \textit{surprise} and \textit{novelty} (overall creativity score: 4.875), likely influenced by participants’ exposure to AI-driven design in Task 2. Of the 24 trials in Task 3, only 4 (17\%) involved perspective changes, as participants preferred adjusting overall video composition over specific objects or perspectives.

In interviews, participants strongly preferred co-creation over passively providing prompts and waiting for results. Half (4/8) favored immersive generation over traditional 2D, describing it as a more complete and unique experience difficult to replicate in daily life. 37.5\% (3/8) indicated they would adopt immersive generation more readily if hardware were more accessible. Challenges with the VR setup included low headset resolution, which made videos appear blurry, and the headset’s bulkiness, which detracted from the immersive experience. Participants also criticized the generative AI model for being overly constrained by image prompts, only allowing camera movements without adding new objects. A film industry participant (Female, 24) expressed concerns about AI replacing human roles in creative fields.

\section{Discussion and Future Work}
The pilot study provided valuable insights into improving the system and advancing collaboration between human and agent in immersive environments.

\textbf{Enhancing Video Generation Models: }
Key improvements to the video generation model are needed to establish a stronger technical foundation for applications. Enhancing video resolution by incorporating advanced architectures, such as vision transformers instead of AnimateDiff, could significantly improve quality, though at the cost of greater computational demands. Additionally, optimizing panoramic video generation by refining the 360-Adapter would ensure seamless outputs with continuous left and right edges and motion patterns that follow curved trajectories. To overcome the current reliance on image prompts and the limited ability to predict diverse motion trajectories, integrating text-to-video and text-to-image models in a collaborative framework could foster the creation of more varied and creative outputs.

\textbf{Improving System Interaction and Usability}: 
Improvements in system interaction and usability are also critical. Strengthening the inference pipeline to better align with Unity applications, such as enabling automated angle adjustments based on user instructions, would improve responsiveness. Moreover, first-time users would benefit from a more intuitive onboarding process, including guided instructions and example showcases to demonstrate the system’s capabilities.

Future work will focus on addressing these technical and interaction challenges, bridging the gap between immersive environments and creative workflows.

\section{Conclusion}

We proposed Imagine360, a proof-of-concept prototype leveraging advancements in video panorama generation, enabling users to interactively create and control panoramic videos through real-time collaboration with AI agents. A pilot study demonstrated the advantages of human-AI collaboration over linear AI-driven design and provided insights for system refinement. Future work will focus on enhancing model capabilities, optimizing interaction strategies, and enabling seamless control of temporal and spatial dimensions to unlock the potential for truly immersive and self-curated realities in virtual environments.

\bibliographystyle{ACM-Reference-Format}
\bibliography{bibfile}


\begin{thebibliography}{34}


\ifx \showCODEN    \undefined \def \showCODEN     #1{\unskip}     \fi
\ifx \showISBNx    \undefined \def \showISBNx     #1{\unskip}     \fi
\ifx \showISBNxiii \undefined \def \showISBNxiii  #1{\unskip}     \fi
\ifx \showISSN     \undefined \def \showISSN      #1{\unskip}     \fi
\ifx \showLCCN     \undefined \def \showLCCN      #1{\unskip}     \fi
\ifx \shownote     \undefined \def \shownote      #1{#1}          \fi
\ifx \showarticletitle \undefined \def \showarticletitle #1{#1}   \fi
\ifx \showURL      \undefined \def \showURL       {\relax}        \fi
\providecommand\bibfield[2]{#2}
\providecommand\bibinfo[2]{#2}
\providecommand\natexlab[1]{#1}
\providecommand\showeprint[2][]{arXiv:#2}

\bibitem[sf_(1999)]%
        {sf_wachowskis1999matrix}
 \bibinfo{year}{1999}\natexlab{}.
\newblock \bibinfo{title}{The Matrix}.
\newblock
\newblock
\shownote{IMDb: tt0133093}.


\bibitem[sf_(2010)]%
        {sf_nolan2010inception}
 \bibinfo{year}{2010}\natexlab{}.
\newblock \bibinfo{title}{Inception}.
\newblock
\newblock
\shownote{IMDb: tt1375666}.


\bibitem[Akimoto et~al\mbox{.}(2022)]%
        {t2i_akimoto2022diverse}
\bibfield{author}{\bibinfo{person}{Naofumi Akimoto}, \bibinfo{person}{Yuhi Matsuo}, {and} \bibinfo{person}{Yoshimitsu Aoki}.} \bibinfo{year}{2022}\natexlab{}.
\newblock \showarticletitle{Diverse plausible 360-degree image outpainting for efficient 3DCG background creation}. In \bibinfo{booktitle}{\emph{Proceedings of the IEEE/CVF Conference on Computer Vision and Pattern Recognition}}. \bibinfo{pages}{11441--11450}.
\newblock


\bibitem[An et~al\mbox{.}(2023)]%
        {latentshift}
\bibfield{author}{\bibinfo{person}{Jie An}, \bibinfo{person}{Songyang Zhang}, \bibinfo{person}{Harry Yang}, \bibinfo{person}{Sonal Gupta}, \bibinfo{person}{Jia-Bin Huang}, \bibinfo{person}{Jiebo Luo}, {and} \bibinfo{person}{Xi Yin}.} \bibinfo{year}{2023}\natexlab{}.
\newblock \showarticletitle{Latent-Shift: Latent Diffusion with Temporal Shift for Efficient Text-to-Video Generation}.
\newblock \bibinfo{journal}{\emph{arXiv preprint arXiv:2304.08477}} (\bibinfo{year}{2023}).
\newblock


\bibitem[Boden(2004)]%
        {boden2004creative}
\bibfield{author}{\bibinfo{person}{Margaret~A. Boden}.} \bibinfo{year}{2004}\natexlab{}.
\newblock \bibinfo{booktitle}{\emph{The Creative Mind: Myths and Mechanisms}}.
\newblock \bibinfo{publisher}{Psychology Press}.
\newblock


\bibitem[Cai et~al\mbox{.}(2024)]%
        {interact360}
\bibfield{author}{\bibinfo{person}{Zeyu Cai}, \bibinfo{person}{Zhelong Huang}, \bibinfo{person}{Xu Zheng}, \bibinfo{person}{Yexin Liu}, \bibinfo{person}{Chao Liu}, \bibinfo{person}{Zeyu Wang}, {and} \bibinfo{person}{Lin Wang}.} \bibinfo{year}{2024}\natexlab{}.
\newblock \showarticletitle{Interact360: Interactive Identity-driven Text to 360° Panorama Generation}. In \bibinfo{booktitle}{\emph{2024 IEEE Conference on Artificial Intelligence (CAI)}}. \bibinfo{pages}{728--736}.
\newblock
\href{https://doi.org/10.1109/CAI59869.2024.00141}{doi:\nolinkurl{10.1109/CAI59869.2024.00141}}


\bibitem[Cheng(2014)]%
        {butterfly}
\bibfield{author}{\bibinfo{person}{Kai-Yuan Cheng}.} \bibinfo{year}{2014}\natexlab{}.
\newblock \showarticletitle{Self and the Dream of the Butterfly in the Zhuangzi}.
\newblock \bibinfo{journal}{\emph{Philosophy East and West}}  \bibinfo{volume}{64} (\bibinfo{date}{07} \bibinfo{year}{2014}), \bibinfo{pages}{563--597}.
\newblock
\href{https://doi.org/10.1353/pew.2014.0051}{doi:\nolinkurl{10.1353/pew.2014.0051}}


\bibitem[Cheng et~al\mbox{.}(2023)]%
        {ddpm_cheng2023null}
\bibfield{author}{\bibinfo{person}{Xinhua Cheng}, \bibinfo{person}{Nan Zhang}, \bibinfo{person}{Jiwen Yu}, \bibinfo{person}{Yinhuai Wang}, \bibinfo{person}{Ge Li}, {and} \bibinfo{person}{Jian Zhang}.} \bibinfo{year}{2023}\natexlab{}.
\newblock \showarticletitle{Null-space diffusion sampling for zero-shot point cloud completion}. In \bibinfo{booktitle}{\emph{Proceedings of the Thirty-Second International Joint Conference on Artificial Intelligence (IJCAI)}}.
\newblock


\bibitem[Chilton et~al\mbox{.}(2021)]%
        {chilton2021visifit}
\bibfield{author}{\bibinfo{person}{Lydia~B Chilton}, \bibinfo{person}{Ecenaz~Jen Ozmen}, \bibinfo{person}{Sam~H Ross}, {and} \bibinfo{person}{Vivian Liu}.} \bibinfo{year}{2021}\natexlab{}.
\newblock \showarticletitle{VisiFit: Structuring Iterative Improvement for Novice Designers}. In \bibinfo{booktitle}{\emph{Proceedings of the 2021 CHI Conference on Human Factors in Computing Systems (CHI '21)}} (Yokohama, Japan). \bibinfo{publisher}{Association for Computing Machinery}, \bibinfo{address}{New York, NY, USA}, Article \bibinfo{articleno}{574}, \bibinfo{numpages}{14}~pages.
\newblock
\href{https://doi.org/10.1145/3411764.3445089}{doi:\nolinkurl{10.1145/3411764.3445089}}


\bibitem[Chung et~al\mbox{.}(2022)]%
        {chung2022talebrush}
\bibfield{author}{\bibinfo{person}{John Joon~Young Chung}, \bibinfo{person}{Wooseok Kim}, \bibinfo{person}{Kang~Min Yoo}, \bibinfo{person}{Hwaran Lee}, \bibinfo{person}{Eytan Adar}, {and} \bibinfo{person}{Minsuk Chang}.} \bibinfo{year}{2022}\natexlab{}.
\newblock \showarticletitle{TaleBrush: Visual Sketching of Story Generation with Pretrained Language Models}. In \bibinfo{booktitle}{\emph{Extended Abstracts of the 2022 CHI Conference on Human Factors in Computing Systems (CHI EA '22)}} (New Orleans, LA, USA). \bibinfo{publisher}{Association for Computing Machinery}, \bibinfo{address}{New York, NY, USA}, Article \bibinfo{articleno}{172}, \bibinfo{numpages}{4}~pages.
\newblock
\href{https://doi.org/10.1145/3491101.3519873}{doi:\nolinkurl{10.1145/3491101.3519873}}


\bibitem[Clark et~al\mbox{.}(2018)]%
        {machineintheloop}
\bibfield{author}{\bibinfo{person}{Elizabeth Clark}, \bibinfo{person}{Anne~Spencer Ross}, \bibinfo{person}{Chenhao Tan}, \bibinfo{person}{Yangfeng Ji}, {and} \bibinfo{person}{Noah~A. Smith}.} \bibinfo{year}{2018}\natexlab{}.
\newblock \showarticletitle{Creative Writing with a Machine in the Loop: Case Studies on Slogans and Stories}. In \bibinfo{booktitle}{\emph{Proceedings of the 23rd International Conference on Intelligent User Interfaces (IUI '18)}} (Tokyo, Japan). \bibinfo{publisher}{ACM}, \bibinfo{address}{New York, NY, USA}, \bibinfo{pages}{329--340}.
\newblock
\href{https://doi.org/10.1145/3172944.3172983}{doi:\nolinkurl{10.1145/3172944.3172983}}


\bibitem[Dalal et~al\mbox{.}(2024)]%
        {dalal2024gaussian}
\bibfield{author}{\bibinfo{person}{Anurag Dalal}, \bibinfo{person}{Daniel Hagen}, \bibinfo{person}{Kjell~G. Robbersmyr}, {and} \bibinfo{person}{Kristian~Muri Knausg{\aa}rd}.} \bibinfo{year}{2024}\natexlab{}.
\newblock \showarticletitle{Gaussian Splatting: 3D Reconstruction and Novel View Synthesis, a Review}.
\newblock \bibinfo{journal}{\emph{arXiv preprint arXiv:2405.03417}} (\bibinfo{year}{2024}).
\newblock


\bibitem[Gero et~al\mbox{.}(2022)]%
        {gero2022sparks}
\bibfield{author}{\bibinfo{person}{Katy~Ilonka Gero}, \bibinfo{person}{Vivian Liu}, {and} \bibinfo{person}{Lydia Chilton}.} \bibinfo{year}{2022}\natexlab{}.
\newblock \showarticletitle{Sparks: Inspiration for Science Writing Using Language Models}. In \bibinfo{booktitle}{\emph{Proceedings of the 2022 ACM Designing Interactive Systems Conference (DIS '22)}} (Virtual Event, Australia). \bibinfo{publisher}{Association for Computing Machinery}, \bibinfo{address}{New York, NY, USA}, \bibinfo{pages}{1002--1019}.
\newblock
\href{https://doi.org/10.1145/3532106.3533533}{doi:\nolinkurl{10.1145/3532106.3533533}}


\bibitem[Gibson(1984)]%
        {sf_gibson1984neuromancer}
\bibfield{author}{\bibinfo{person}{William Gibson}.} \bibinfo{year}{1984}\natexlab{}.
\newblock \bibinfo{booktitle}{\emph{Neuromancer}}.
\newblock \bibinfo{publisher}{Ace Books}, \bibinfo{address}{New York}.
\newblock
\newblock
\shownote{ISBN: 978-0441569595}.


\bibitem[Gmeiner et~al\mbox{.}(2023)]%
        {gmeiner2023exploring}
\bibfield{author}{\bibinfo{person}{Frederic Gmeiner}, \bibinfo{person}{Humphrey Yang}, \bibinfo{person}{Lining Yao}, \bibinfo{person}{Kenneth Holstein}, {and} \bibinfo{person}{Nikolas Martelaro}.} \bibinfo{year}{2023}\natexlab{}.
\newblock \showarticletitle{Exploring Challenges and Opportunities to Support Designers in Learning to Co-Create with AI-Based Manufacturing Design Tools}. In \bibinfo{booktitle}{\emph{Proceedings of the 2023 CHI Conference on Human Factors in Computing Systems (CHI '23)}} (Hamburg, Germany). \bibinfo{publisher}{Association for Computing Machinery}, \bibinfo{address}{New York, NY, USA}, Article \bibinfo{articleno}{226}, \bibinfo{numpages}{20}~pages.
\newblock
\href{https://doi.org/10.1145/3544548.3580999}{doi:\nolinkurl{10.1145/3544548.3580999}}


\bibitem[Guo et~al\mbox{.}(2023)]%
        {animatediff}
\bibfield{author}{\bibinfo{person}{Yuwei Guo}, \bibinfo{person}{Ceyuan Yang}, \bibinfo{person}{Anyi Rao}, \bibinfo{person}{Zhengyang Liang}, \bibinfo{person}{Yaohui Wang}, \bibinfo{person}{Yu Qiao}, \bibinfo{person}{Maneesh Agrawala}, \bibinfo{person}{Dahua Lin}, {and} \bibinfo{person}{Bo Dai}.} \bibinfo{year}{2023}\natexlab{}.
\newblock \bibinfo{title}{AnimateDiff: Animate Your Personalized Text-to-Image Diffusion Models without Specific Tuning}.
\newblock


\bibitem[Hart and Staveland(1988)]%
        {tlx}
\bibfield{author}{\bibinfo{person}{Sandra~G Hart} {and} \bibinfo{person}{Lowell~E Staveland}.} \bibinfo{year}{1988}\natexlab{}.
\newblock \showarticletitle{Development of NASA-TLX (Task Load Index): Results of empirical and theoretical research}.
\newblock In \bibinfo{booktitle}{\emph{Human mental workload}}. \bibinfo{publisher}{Elsevier}, \bibinfo{pages}{139--183}.
\newblock


\bibitem[Ho et~al\mbox{.}(2020)]%
        {ddpm_ho2020denoising}
\bibfield{author}{\bibinfo{person}{Jonathan Ho}, \bibinfo{person}{Ajay Jain}, {and} \bibinfo{person}{Pieter Abbeel}.} \bibinfo{year}{2020}\natexlab{}.
\newblock \showarticletitle{Denoising diffusion probabilistic models}.
\newblock \bibinfo{journal}{\emph{Advances in Neural Information Processing Systems}}  \bibinfo{volume}{33} (\bibinfo{year}{2020}), \bibinfo{pages}{6840--6851}.
\newblock


\bibitem[Ibarrola et~al\mbox{.}(2023)]%
        {ibarrola2023collaborative}
\bibfield{author}{\bibinfo{person}{Francisco Ibarrola}, \bibinfo{person}{Tomas Lawton}, {and} \bibinfo{person}{Kazjon Grace}.} \bibinfo{year}{2023}\natexlab{}.
\newblock \showarticletitle{A Collaborative, Interactive and Context-Aware Drawing Agent for Co-Creative Design}.
\newblock \bibinfo{journal}{\emph{IEEE Transactions on Visualization and Computer Graphics}} (\bibinfo{year}{2023}), \bibinfo{pages}{1--13}.
\newblock
\href{https://doi.org/10.1109/TVCG.2023.3293853}{doi:\nolinkurl{10.1109/TVCG.2023.3293853}}


\bibitem[Kant(1790)]%
        {kant1790critique}
\bibfield{author}{\bibinfo{person}{Immanuel Kant}.} \bibinfo{year}{1790}\natexlab{}.
\newblock \bibinfo{booktitle}{\emph{Critique of Judgment}}.
\newblock \bibinfo{publisher}{Penguin Classics}, \bibinfo{address}{London}.
\newblock
\newblock
\shownote{Original work published in 1790}.


\bibitem[Kantosalo and Toivonen(2016)]%
        {kantosalo2016creative}
\bibfield{author}{\bibinfo{person}{Anna Kantosalo} {and} \bibinfo{person}{Hannu Toivonen}.} \bibinfo{year}{2016}\natexlab{}.
\newblock \showarticletitle{Modes for creative human-computer collaboration: Alternating and task-divided co-creativity}. In \bibinfo{booktitle}{\emph{Proceedings of the Seventh International Conference on Computational Creativity}}. \bibinfo{pages}{77--84}.
\newblock


\bibitem[Lan et~al\mbox{.}(2024)]%
        {immersivesketch}
\bibfield{author}{\bibinfo{person}{Alfred Lan}, \bibinfo{person}{Tai-Chen Tsai}, \bibinfo{person}{Chih-Chuan Huang}, \bibinfo{person}{Pu Ching}, \bibinfo{person}{Tse-Yu Pan}, {and} \bibinfo{person}{Min-Chun Hu}.} \bibinfo{year}{2024}\natexlab{}.
\newblock \showarticletitle{ImmerseSketch: Transforming Creative Prompts into Vivid 3D Environments in VR}. In \bibinfo{booktitle}{\emph{ACM SIGGRAPH 2024 Posters}} (Denver, CO, USA) \emph{(\bibinfo{series}{SIGGRAPH '24})}. \bibinfo{publisher}{Association for Computing Machinery}, \bibinfo{address}{New York, NY, USA}, Article \bibinfo{articleno}{56}, \bibinfo{numpages}{2}~pages.
\newblock
\showISBNx{9798400705168}
\href{https://doi.org/10.1145/3641234.3671078}{doi:\nolinkurl{10.1145/3641234.3671078}}


\bibitem[Mou et~al\mbox{.}(2024)]%
        {t2i_mou2024t2i}
\bibfield{author}{\bibinfo{person}{Chong Mou}, \bibinfo{person}{Xintao Wang}, \bibinfo{person}{Liangbin Xie}, \bibinfo{person}{Yanze Wu}, \bibinfo{person}{Jian Zhang}, \bibinfo{person}{Zhongang Qi}, {and} \bibinfo{person}{Ying Shan}.} \bibinfo{year}{2024}\natexlab{}.
\newblock \showarticletitle{T2i-adapter: Learning adapters to dig out more controllable ability for text-to-image diffusion models}. In \bibinfo{booktitle}{\emph{Proceedings of the AAAI Conference on Artificial Intelligence}}. \bibinfo{pages}{4296--4304}.
\newblock


\bibitem[Nichol et~al\mbox{.}(2021)]%
        {t2i_nichol2021glide}
\bibfield{author}{\bibinfo{person}{Alex Nichol}, \bibinfo{person}{Prafulla Dhariwal}, \bibinfo{person}{Aditya Ramesh}, \bibinfo{person}{Pranav Shyam}, \bibinfo{person}{Pamela Mishkin}, \bibinfo{person}{Bob McGrew}, \bibinfo{person}{Ilya Sutskever}, {and} \bibinfo{person}{Mark Chen}.} \bibinfo{year}{2021}\natexlab{}.
\newblock \showarticletitle{GLIDE: Towards photorealistic image generation and editing with text-guided diffusion models}.
\newblock \bibinfo{journal}{\emph{arXiv preprint arXiv:2112.10741}} (\bibinfo{year}{2021}).
\newblock


\bibitem[Schopenhauer(1818)]%
        {schopenhauer1818world}
\bibfield{author}{\bibinfo{person}{Arthur Schopenhauer}.} \bibinfo{year}{1818}\natexlab{}.
\newblock \bibinfo{booktitle}{\emph{The World as Will and Representation}}.
\newblock \bibinfo{publisher}{Dover Publications}, \bibinfo{address}{New York}.
\newblock
\newblock
\shownote{Original work published in 1818}.


\bibitem[Singer et~al\mbox{.}(2022)]%
        {makeavideo}
\bibfield{author}{\bibinfo{person}{Uriel Singer}, \bibinfo{person}{Adam Polyak}, \bibinfo{person}{Thomas Hayes}, \bibinfo{person}{Xi Yin}, \bibinfo{person}{Jie An}, \bibinfo{person}{Songyang Zhang}, \bibinfo{person}{Qiyuan Hu}, \bibinfo{person}{Harry Yang}, \bibinfo{person}{Oron Ashual}, \bibinfo{person}{Oran Gafni}, {et~al\mbox{.}}} \bibinfo{year}{2022}\natexlab{}.
\newblock \showarticletitle{Make-a-video: Text-to-video generation without text-video data}.
\newblock \bibinfo{journal}{\emph{arXiv preprint arXiv:2209.14792}} (\bibinfo{year}{2022}).
\newblock


\bibitem[Stephenson(1992)]%
        {sf_stephenson1992snowcrash}
\bibfield{author}{\bibinfo{person}{Neal Stephenson}.} \bibinfo{year}{1992}\natexlab{}.
\newblock \bibinfo{booktitle}{\emph{Snow Crash}}.
\newblock \bibinfo{publisher}{Bantam Books}, \bibinfo{address}{New York}.
\newblock
\newblock
\shownote{ISBN: 978-0553380958}.


\bibitem[Tochilkin et~al\mbox{.}(2024)]%
        {tochilkin2024triposr}
\bibfield{author}{\bibinfo{person}{Dmitry Tochilkin}, \bibinfo{person}{David Pankratz}, \bibinfo{person}{Zexiang Liu}, \bibinfo{person}{Zixuan Huang}, \bibinfo{person}{Adam Letts}, \bibinfo{person}{Yangguang Li}, \bibinfo{person}{Ding Liang}, \bibinfo{person}{Christian Laforte}, \bibinfo{person}{Varun Jampani}, {and} \bibinfo{person}{Yan-Pei Cao}.} \bibinfo{year}{2024}\natexlab{}.
\newblock \showarticletitle{{TripoSR}: Fast 3D Object Reconstruction from a Single Image}.
\newblock \bibinfo{journal}{\emph{arXiv preprint arXiv:2403.02151}} (\bibinfo{year}{2024}).
\newblock


\bibitem[Wang et~al\mbox{.}(2024b)]%
        {panoramat2i}
\bibfield{author}{\bibinfo{person}{Hai Wang}, \bibinfo{person}{Xiaoyu Xiang}, \bibinfo{person}{Yuchen Fan}, {and} \bibinfo{person}{Jing-Hao Xue}.} \bibinfo{year}{2024}\natexlab{b}.
\newblock \showarticletitle{{ Customizing 360-Degree Panoramas through Text-to-Image Diffusion Models }}. In \bibinfo{booktitle}{\emph{2024 IEEE/CVF Winter Conference on Applications of Computer Vision (WACV)}}. \bibinfo{publisher}{IEEE Computer Society}, \bibinfo{address}{Los Alamitos, CA, USA}, \bibinfo{pages}{4921--4931}.
\newblock
\href{https://doi.org/10.1109/WACV57701.2024.00486}{doi:\nolinkurl{10.1109/WACV57701.2024.00486}}


\bibitem[Wang et~al\mbox{.}(2024a)]%
        {360dvd}
\bibfield{author}{\bibinfo{person}{Qian Wang}, \bibinfo{person}{Weiqi Li}, \bibinfo{person}{Chong Mou}, \bibinfo{person}{Xinhua Cheng}, {and} \bibinfo{person}{Jian Zhang}.} \bibinfo{year}{2024}\natexlab{a}.
\newblock \showarticletitle{360DVD: Controllable Panorama Video Generation with 360-Degree Video Diffusion Model}.
\newblock \bibinfo{journal}{\emph{arXiv preprint arXiv:2401.06578}} (\bibinfo{year}{2024}).
\newblock


\bibitem[Yankelevich and Zaragoza(2014)]%
        {apprentice}
\bibfield{author}{\bibinfo{person}{Santiago~Negrete Yankelevich} {and} \bibinfo{person}{Nora Angelica~Morales Zaragoza}.} \bibinfo{year}{2014}\natexlab{}.
\newblock \showarticletitle{The apprentice framework: planning and assessing creativity}. In \bibinfo{booktitle}{\emph{Proceedings of the International Conference on Computational Creativity}} (Jönköping, Sweden). \bibinfo{publisher}{Association for Computational Creativity}, \bibinfo{pages}{280--283}.
\newblock


\bibitem[Yu et~al\mbox{.}(2024)]%
        {vrpainting}
\bibfield{author}{\bibinfo{person}{Emilie Yu}, \bibinfo{person}{Fanny Chevalier}, \bibinfo{person}{Karan Singh}, {and} \bibinfo{person}{Adrien Bousseau}.} \bibinfo{year}{2024}\natexlab{}.
\newblock \showarticletitle{3D-Layers: Bringing Layer-Based Color Editing to VR Painting}.
\newblock  \bibinfo{volume}{43}, \bibinfo{number}{4}, Article \bibinfo{articleno}{101} (\bibinfo{date}{July} \bibinfo{year}{2024}), \bibinfo{numpages}{15}~pages.
\newblock
\showISSN{0730-0301}
\href{https://doi.org/10.1145/3658183}{doi:\nolinkurl{10.1145/3658183}}


\bibitem[Yuan et~al\mbox{.}(2024)]%
        {vrpaintingfinger}
\bibfield{author}{\bibinfo{person}{Rosina Yuan}, \bibinfo{person}{Antony Tang}, \bibinfo{person}{Qianyuan Zou}, \bibinfo{person}{Masoumeh~Hesam Mahmoudinezhad}, \bibinfo{person}{Yuewei Zhang}, {and} \bibinfo{person}{Iain Anderson}.} \bibinfo{year}{2024}\natexlab{}.
\newblock \showarticletitle{Finger Painting in VR: Multi-Dynamic Gestural Input for VR Painting}. In \bibinfo{booktitle}{\emph{SIGGRAPH Asia 2024 XR}} \emph{(\bibinfo{series}{SA '24})}. \bibinfo{publisher}{Association for Computing Machinery}, \bibinfo{address}{New York, NY, USA}, Article \bibinfo{articleno}{6}, \bibinfo{numpages}{2}~pages.
\newblock
\showISBNx{9798400711411}
\href{https://doi.org/10.1145/3681759.3688918}{doi:\nolinkurl{10.1145/3681759.3688918}}


\bibitem[Zhou et~al\mbox{.}(2022)]%
        {magicvideo}
\bibfield{author}{\bibinfo{person}{Daquan Zhou}, \bibinfo{person}{Weimin Wang}, \bibinfo{person}{Hanshu Yan}, \bibinfo{person}{Weiwei Lv}, \bibinfo{person}{Yizhe Zhu}, {and} \bibinfo{person}{Jiashi Feng}.} \bibinfo{year}{2022}\natexlab{}.
\newblock \showarticletitle{Magicvideo: Efficient video generation with latent diffusion models}.
\newblock \bibinfo{journal}{\emph{arXiv preprint arXiv:2211.11018}} (\bibinfo{year}{2022}).
\newblock


\end{thebibliography}

\end{document}